# Mineral Processing and Metal Extraction on the Lunar Surface - Challenges and Opportunities

Matthew G. Shaw[1]*, Matthew S. Humbert[1], Geoffrey A. Brooks[1], Akbar Rhamdhani[1], Alan R. Duffy[2], and Mark I. Pownceby[3]

1. Fluid and Process Dynamics Group, FSET, Swinburne University of Technology, Hawthorn, VIC 3122, Australia

2. Centre for Astrophysics and Supercomputing, Swinburne University of Technology, Hawthorn, VIC 3122, Australia

3. CSIRO Mineral Resources, Clayton, VIC 3168, Australia

* Corresponding author: mgshaw@swin.edu.au

**Abstract:** The lunar surface is extremely harsh and current mineral processing and metal extraction technologies are not adequately equipped to address this environment. In this paper we review the metals available for extraction and conditions at the lunar surface, and analyse the challenges associated with comminution, beneficiation, and metal extraction operations. The potential beneficial effects of the natural lunar conditions are also evaluated. This investigation concludes that process plant design on the lunar surface will favour lightweight, schematically simple flow sheets that enable automation, and that utilise the local environment wherever possible. The elimination of traditional comminution and beneficiation stages and their replacement with basic classification could be economically favourable. The most promising metal reduction pathways are identified as molten regolith electrolysis, and vacuum thermal dissociation, other processes with merit are hydrogen reduction, carbothermal reduction, and solid electrolysis. Finally, it is identified that a significant research effort in all areas of astrometallurgy will be required before industrial-sized extra-terrestrial mineral processing and metal extraction operations will be viable.

# Introduction

The development of metallurgical processes for use in space, termed astrometallurgy, is a field of research that will be critical for prolonged human and robotic presence in cislunar space and beyond. The processing and use of in situ resources, or space resource utilisation (SRU), is intended to enable the economic industrialisation of space, and specifically in the case of the current work, the Moon. With a prolonged presence on the Moon will come the need for more substantial and complex mineral extraction and metal production operations.

Mineral processing or 'beneficiation', as a practice, is predicated on the idea of separating valuable minerals from gangue (waste) based on their different physical and chemical characteristics. The extent to which those characteristics can be exploited is heavily influenced by the surrounding conditions: gravity, pressure, exposure to water, etc. These factors will become major considerations in the design and use of metallurgical processing technologies in space. The ambient conditions on the surface of the Earth have inherently led to the development of existing mineral processing and metal extraction technologies that take advantage of these conditions [1-3]. In some processes, such as pressure leaching or separation via cyclones, the prevailing conditions are enhanced, usually through *increasing* pressure or gravity to enhance separation. Conditions more relevant to processing at the lunar surface, such as the *reduction* of pressure or gravity, are rarely if ever considered in process plant design. In addition, most Earth-based mineral separation processes use water as a fluidising medium to effect separation and, whilst water management in a processing plant can consume significant resources, there is always a fundamental assumption that water will be available for use. The ready supply of water will almost certainly not be the case in off-Earth environments. When considering the development of mineral processing and metal extraction technologies for use in space, these accepted assumptions are of critical importance. The challenging reality is that the majority of processing technologies that have been refined over hundreds of years for use on Earth may not work in space. This is not to say that all current technologies are useless, some may be able to be adapted with only minor modifications but the question must be raised of whether the modification of technologies designed for use on Earth will result in the most efficient processing technologies. In the words of Haskin [4] when considering lunar processing, "Processes that take advantage of the lunar environment deserve at least equal attention".

A significant amount of work has been invested over the last 60 years into the conceptualisation and demonstration of resource extraction from lunar minerals. This work has predominantly focused on the production of oxygen [5, 6], however, the production of oxygen from a metal oxide often results in a metallic by-product and thus is of interest in the form of a potential metallurgical extraction technique. On that note, while the target resource of traditional extractive metallurgy has always been metals, in the case of astrometallurgical processing, oxygen recovery will also play a significant role in process economic viability. Finally, the resources extracted in the methods presented in this review are aimed at use in space. While for terrestrially rarer resources it may, in the future, become economically viable to transport them back down to Earth, the costs of transport are currently too high. The current work will assume all extracted resources are intended for use in space-based activities.

The development of robust mineral processing and metal extraction processes and equipment for use on the Moon will require a thorough understanding of how the local environment will affect existing processes. In some cases, existing processes will need only minor modifications for conversion for In Situ Resource Utilisation (ISRU) activities; in others, entirely new processes will be required to perform tasks that have been rendered uneconomical by the local environment. The current work, therefore, focusses on the broader problem of processing on the lunar surface and the effects, both beneficial and detrimental, of the lunar environmental conditions on potential mineral extraction and metal reduction technologies. The oxidised metal feedstocks available on the lunar surface will be described, along with the characteristics of their occurrence: grain size, geochemical and modal composition, and general petrography. The ambient conditions on the Moon will be contrasted to those of Earth and the effects on metallurgical equipment and processes will be discussed. Historically proposed metal extraction processes will be analysed for viability in industrial sized processing operations. Finally, the potential beneficial effects of the 'challenging' lunar environment will be discussed in terms of possible avenues for the development of novel technologies optimised for use in space. It will be shown in the current work that there is a large amount of research that will be required in the field of astrometallurgy before industrially sized mineral processing and metal reduction processes operating on off-Earth bodies can be considered viable options for a source of metal beyond low Earth orbit.

## Resource availability

When designing an extraction process, knowledge of the specific chemical and mineralogical compositions, and the physical properties of a process feedstock are important. Compared to the level of detail usually required for process design on Earth, there is very limited data available as to the composition of the Moon. Though knowledge of lunar geology is restricted to the samples returned on the Apollo and Luna missions and from orbital mapping by satellites, there is a general consensus that the lunar geology is significantly more homogenous that that found on Earth [7, 8]. This homogeneity is attributed to the lack of surface altering processes on the Moon. On Earth, plate tectonics, active volcanism, and weathering play significant roles in the geochemical and mineralogical evolution and composition of the feed materials used in mineral processing and metal extraction technologies. Unlike Earth, the predominant surface altering process on the Moon is that of meteoric and micro-meteoric impacts [9]. This natural comminution of the lunar surface results in two main potential feed sources for industrial mineral processing and metal extraction processes, the bedrock, and the fine-grained material formed by this natural weathering, the regolith. In the current work we will look at both potential feedstocks, the bedrock and the regolith, however, more focus will be given to the regolith material as it is a more viable feedstock as will be discussed here.

Petrographically, the Moon is divided into two major regions, the lighter coloured Al-rich Highlands or Terra, and the darker Fe- and Ti-rich Maria or Mare. Figure 1 clearly shows these two major petrographic regions on the near side of the Moon. The majority of the far side of the Moon is characterised by comprising more Highlands type material. For more detailed mapping based on Clementine data [10] see the ternary element diagrams presented by Spudis *et al.* [11].

*Lunar rocks*

The composition of ten returned lunar rock samples, representing the presumed composition of the bedrock based on Apollo mission samples, are provided in Table 1. The data, in mass percent, are split into Mare and Highlands compositions and represent examples of the main rock types encountered during lunar exploration to date [12]. Further detail on these rock types and detailed mineralogical and geochemical analysis can be found elsewhere [8, 13-16].

Highland bedrock geology appears to vary somewhat with a large fraction being predominantly anorthositic in nature. In the Maria, there are two main categories, the High-Ti, and Low-Ti maria [12]. Mapping of the bedrock compositions on the Moon is difficult as most areas are covered in a layer of regolith material. The comminution mechanism that forms the regolith also results in the mixing of the regolith from different petrographic areas, as such, the composition of the regolith tends to be more homogenous than that of the underlying bedrock.

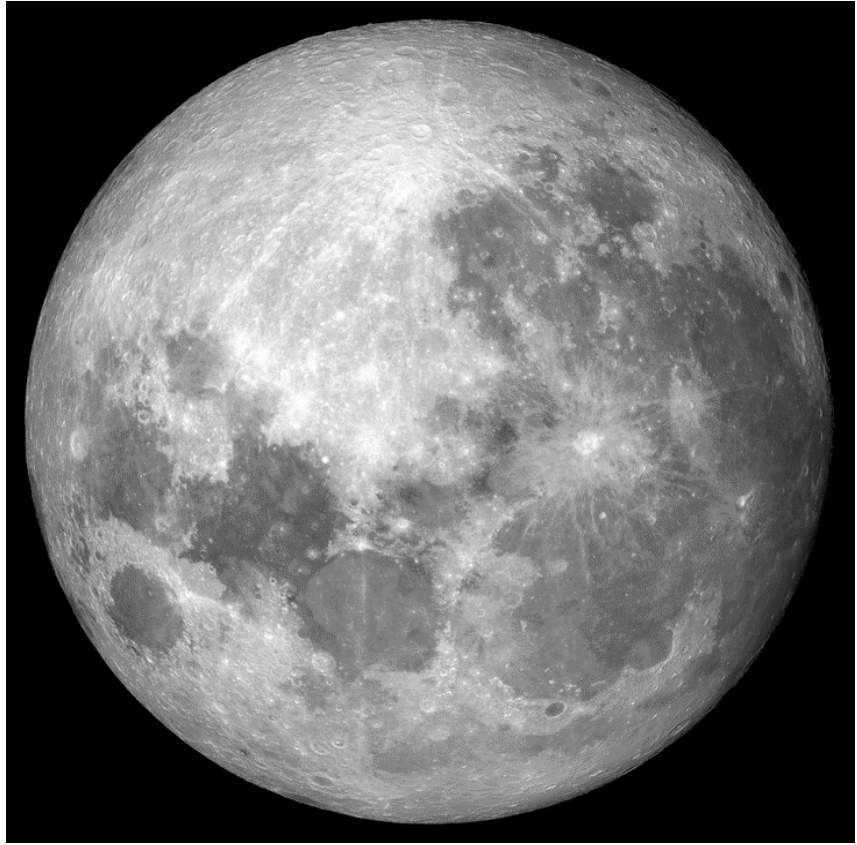

*Figure 1 - Black and white image of the Moon showing the major petrographic regions. The lighter coloured Al-rich Highlands, and the darker more Ti- and Fe-rich Mare. (Image: NASA).*

*Table 1 - Common lunar rock chemical compositions (mass %). Table constructed using data summarised from Duke et al. [12].*

|  | **Mare Rocks** | | | | **Highland Rocks** | | | | | |
| --- | --- | --- | --- | --- | --- | --- | --- | --- | --- | --- |
|  | High-Ti | Low-Ti | Very Low-Ti | Al-rich | Anorthosite | Norite | Troctolite | KREEP Basalt* | QMD** | Granite |
| $SiO_2$ | 39.7 | 45.8 | 46 | 46.4 | 45.3 | 51.1 | 42.9 | 50.8 | 56.9 | 74.2 |
| $TiO_2$ | 11.2 | 2.8 | 1.1 | 2.6 | <0.02 | 0.34 | 0.05 | 2.2 | 1.1 | 0.33 |
| $Al_2O_3$ | 9.5 | 9.6 | 12.1 | 13.6 | 34.2 | 15 | 20.7 | 14.8 | 6.4 | 12.5 |
| $Cr_2O_3$ | 0.37 | 0.56 | 0.27 | 0.4 | 0.004 | 0.38 | 0.11 | 0.31 | 0.16 | 0.002 |
| FeO | 19.0 | 20.2 | 22.1 | 16.8 | 0.5 | 10.7 | 5.0 | 10.6 | 18.6 | 2.32 |
| MnO | 0.25 | 0.27 | 0.28 | 0.26 | 0.008 | 0.17 | 0.07 | 0.16 | 0.28 | 0.02 |
| MgO | 7.8 | 9.7 | 6.0 | 8.5 | 0.21 | 12.9 | 19.1 | 8.2 | 4.7 | 0.07 |
| CaO | 11.2 | 10.2 | 11.6 | 11.2 | 19.8 | 8.8 | 11.4 | 9.7 | 8.3 | 1.3 |
| $Na_2O$ | 0.38 | 0.34 | 0.26 | 0.4 | 0.45 | 0.38 | 0.2 | 0.73 | 0.52 | 0.52 |
| $K_2O$ | 0.05 | 0.06 | 0.02 | 0.01 | 0.11 | 0.18 | 0.03 | 0.67 | 2.17 | 8.6 |
| $P_2O_5$ | 0.06 | 0.05 |  |  |  |  | 0.03 | 0.7 | 1.33 |  |
| S | 0.19 | 0.09 |  |  |  |  |  |  |  |  |
| Total | 99.7 | 99.7 | 99.7 | 100.2 | 100.6 | 99.9 | 99.6 | 98.9 | 100.5 | 99.9 |
| Ref. | [13] | [13] | [14] | [14] | [15] | [15] | [15] | [15] | [15] | [15] |

*a feldspathic basalt characterised by elevated levels of Potassium (K), Rare Earth Elements (REE), and Phosphorus (P).
**Quartz Monzodiorite

When dealing with macro concentrations, the composition of the lunar surface is composed primarily of those elements listed in Table 1. In addition to the major elements, there are a number of volatile elements that are known to exist within the regolith and surface lunar rock. These elements are implanted through impact activity and account for less than 1% of the total lunar surface chemistry. The two predominant suspected sources of volatiles are the solar wind and small solar system bodies (meteoroids, comets, and asteroids), however, other origins such as intergalactic dust and endogenous sources also exist [17]. The implanted volatiles are measured in the ppm range and consist among them a number of potentially useful elements such as H, C, N, He, F and Cl. Table 2 shows the average values of these elements in Apollo rock and regolith samples, note that these volatiles exist in significantly higher concentrations in the regolith material as opposed to rock samples, this is presumably due to the higher surface area available for solar wind deposition [18]. Other volatile elements that can be found in ppm and ppb concentrations within the lunar regolith are: He, Ne, Ar, Kr, Xe, Zn, Cd, Au, Ge, Ag, Pb, Tl, S, Sb, Br, and others [17]. These elements will not be considered in the current work, however, there exists the potential for them to be the target of resource extraction in the future.

*Table 2 - Implanted H, C, N, F, Cl and He concentrations of Apollo samples, data by Fegley and Swingle [18].*

| Element | Sample type | Average content (ppm) |
|---|---|---|
| **H** | Regolith | 46 |
|  | Basalt | 2.7 |
| **C** | Regolith | 124 |
|  | Basalt | 26 |
| **N** | Regolith | 81 |
|  | Basalt | 19 |
| **F** | Regolith | 70 |
| **Cl** | Regolith | 30 |
| $^4$**He** | Regolith | 14 |
| $^3$**He** | Regolith | 0.0042 |

As discussed briefly above, due to the more accessible nature of the regolith on the lunar surface, and the costs associated with the transport of hard rock mining equipment from Earth, this material is considered of most interest in terms of lunar SRU feedstocks. In the following section, the properties of the lunar regolith are discussed with reference to how these may be exploited in mineral processing operations.

*The lunar regolith*

The unconsolidated layer of regolith material varies between 2 to 10 m in depth all over the Moon [7-9]. Due to impact weathering, the lunar regolith is very glassy; this glass is often found in the form of breccias, agglutinates, or as minerals embedded in a glass matrix [7, 8, 16]. Impact weathering also results in a lack of smoothing of small particles and rocks that would be expected from terrestrial weathering processes. The high glass content and jagged grain surfaces results in the regolith being a very abrasive substance, the consequences of which cannot be ignored when considering processing equipment wear rates and handling procedures.

The average grain size of the returned lunar regolith samples can be seen in Figure 2. The data used here consists of 127 data sets sorted from that presented by Graf [19]. This data consists of samples from Apollo 12, 15, 16, 17, and Luna 24. The mean P50, or 50% passing size of this data set is 69μm. The P80, more often used for mineral processing analysis, is 257μm. When compared to a standard target size for a ball mill discharge, this is quite course, and there exists significant variability within the regolith in terms of grain sizes. In this data set there were samples that ranged from a P80 of 96μm to 2426μm. The regolith consists of a relatively large fraction of ultra-fine material with extreme cases such as soil #66075,16 which consisted of 41% passing 20 μm [19].

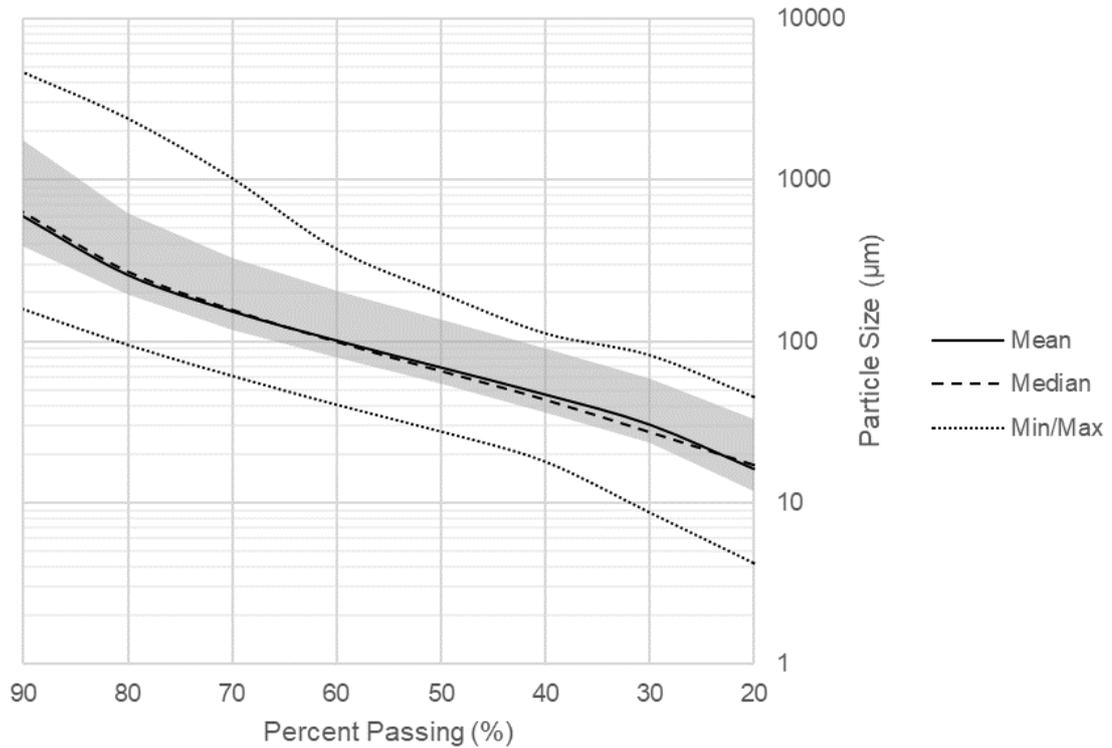

*Figure 2 – Average particle size distribution of Apollo and Luna returned regolith samples, presented in %passing sieve size. Data from Graf [19], sorted to most common sieve set only.*

Table 3 shows the geochemical composition of 9 regolith samples returned on the Apollo and Luna missions in mass percent. These data have been ordered from least FeO to most FeO in the sample roughly equating to more Highlands type material to Maria type respectively. Detailed analysis of these samples can be found elsewhere [16, 20-23]. Until more detailed mapping and sampling campaigns can be completed on the lunar surface providing higher resolution resource maps and confirming or disproving the existence of localised metal concentrations as found on Earth, the regolith compositions displayed here are the presumed range of compositions available for SRU feedstock material on the Moon.

*Table 3 - Returned regolith sample compositions (mass %), ordered by FeO content from left to right. Data from Papike et al. [16]*

| Mission | Apollo 16 | Luna 20 | Apollo 14 | Apollo 17 | Apollo 15 | Apollo 12 | Apollo 11 | Luna 16 | Luna 24 |
|---|---|---|---|---|---|---|---|---|---|
| Sample | 64501 | 22001 | 14163 | 76501 | 15271 | 12033 | 10084 | 21000 | 24999 |
| $SiO_2$ | 45.2 | * | 47.4 | 42.8 | 46.3 | 47.0 | 41.3 | * | * |
| $TiO_2$ | 0.4 | 0.5 | 1.6 | 3.3 | 1.4 | 2.5 | 7.3 | 3.3 | 1.0 |
| $Al_2O_3$ | 27.6 | 22.9 | 17.5 | 18.2 | 16.2 | 13.8 | 13.6 | 14.9 | 11.1 |
| $Cr_2O_3$ | 0.1 | 0.2 | 0.2 | 0.3 | 0.4 | 0.4 | 0.3 | 0.3 | 0.4 |
| FeO | 4.4 | 7.4 | 10.4 | 11.1 | 12.9 | 15.1 | 16.0 | 16.4 | 20.3 |
| MnO | 0.1 | 0.1 | 0.1 | 0.1 | 0.2 | 0.2 | 0.2 | 0.2 | 0.3 |
| MgO | 4.7 | 8.9 | 10.1 | 11.9 | 11.1 | 9.5 | 8.3 | 8.3 | 10.4 |
| CaO | 16.6 | 14.2 | 11.3 | 12.3 | 11.1 | 10.6 | 12.3 | 11.8 | 10.7 |
| $Na_2O$ | 0.4 | 0.3 | 0.7 | 0.4 | 0.5 | 0.7 | 0.4 | 0.4 | 0.3 |
| $K_2O$ | 0.1 | 0.1 | 0.6 | 0.1 | 0.2 | 0.4 | 0.2 | 0.1 | 0.0 |
| Total | 99.6 | * | 99.9 | 100.5 | 100.2 | 100.1 | 99.9 | * | * |

*No Si assays exist from this data set for the Luna samples [20]. For an estimate, the Si content can be calculated as the remainder.

One other metric by which the lunar regolith varies from that found on Earth is the lack of water. This lack of water on the Moon affects the geochemistry and mineralogy of the feed material. Minerals containing ferric iron ($Fe^{3+}$) were noticeably absent in the returned Apollo and Luna samples along with any minerals that contain water (clays, micas, amphiboles, etc.)[7, 8]. While large quantities of water have been confirmed in permanently shadowed regions (PSRs) on the lunar poles [24, 25], and have been proposed to exist in other PSRs even in equatorial regions [26]. No physical samples of these areas have been examined to date. It is unclear how much of this water is in the form of water ice and how much is included within rocks themselves in the form of hydrated minerals. Regardless of the potential existence of hydrated minerals in these PSRs, most of the regolith available on the Moon as feedstock for extractive metallurgical processes will be anhydrous. Water ice mining for life support and rocket fuel will almost certainly be one of the first forms of industrial activity on the lunar surface and these topics have been extensively covered [17, 27-36]. There is an argument to be made that the use of material that has been already processed to remove the water and other volatiles may become an important feedstock for metal extraction processes as this approach will minimise material handling and mining activities, however there is no indication that the base regolith composition from these volatile rich regions varies compared to that presented here.

The bulk modal composition of the Apollo 16 regolith sample #64501 can be seen in Figure 3. This data, presented by Papike *et al.* [16] (1000 to 90 μm) and Labotka *et al.* [23] (90 to 20 μm, and 20 to 10 μm), shows the significant glassy nature of the regolith. Different glass types and mineral types have been combined in this figure for simplicity. The 64501 sample is a good representation of 'clean' highlands material [16], that is, regolith that has been derived from predominantly highlands bedrock with minimal mixing of maria derived components, this can be seen in Table 3 by the low FeO and $TiO_2$ content. The smaller size fractions of this sample are largely composed of glass and monomineralic fragments, whereas the larger size fraction consists of significantly larger amounts of fused soil components and lithic fragments.

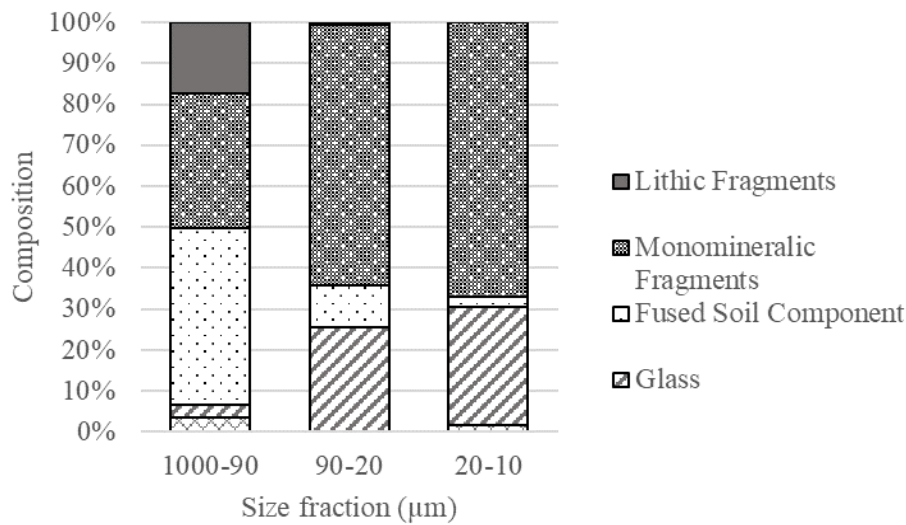

*Figure 3 –Simplified modal composition of the lunar regolith sample #64501 split into three size fractions (1000 to 90 μm, 90 to 20 μm, and 20 to 10 μm), data from Papike et al. [16], and Labotka et al.[23].*

The exact effects on processing technologies of the mineralogical and geochemical differences between terrestrial and lunar feedstocks, the high glass content and lack of hydrated minerals in the lunar regolith, is uncertain. These effects may be minimal, but it is yet again a factor that warrants further investigation.

*Target metals*

Whilst most materials found on the Moon are not in sufficient concentrations to be considered ore on Earth, their mere existence outside Earth's gravity-well, and thus significant reduction in transportation costs for use in space, make them significantly more valuable [35]. As noted previously and contrary to a common misconception, the mining and processing of space-native resources is currently solely targeted towards space-based industry. The transport of resources down to earth in industrial quantities is currently too expensive to be considered as a competitive alternative to terrestrial industry. Initially, space materials processing will be targeted towards materials that will enable lunar colonization and provide propellant to space missions; however, with time, any material that can be efficiently harvested on the Moon will become valuable for use beyond low Earth orbit.

The main metals contained in the lunar regolith along with some of their potential uses can be found in Table 4. It is important to note that along with the impacts on processing equipment that will be covered later in the current work, the effect of gravity and vacuum on construction materials, namely the lowered strength requirements for structural support and the lack of oxidising atmosphere, render metals such as aluminium much more viable for bulk construction than they would normally be considered on Earth. There has also been considerable research into the use of ceramics and waterless concretes using raw, unprocessed regolith; such uses do not require significant processing and are outside the scope of the current work. For a recent review of these potential extra-terrestrial construction materials see Naser [37, 38].

*Table 4 - Metals available for extraction on the lunar surface*

| Metal | Uses | Refs |
|---|---|---|
| Silicon | Electronics and photovoltaic panels, Silanes ($SiH_x$) as rocket fuel alternative. Energy carrier/storage | [39-41] |
| Aluminium | Construction material (pure or alloyed) Solid powder as rocket fuel Energy carrier/storage | [33, 37, 38, 41-43] |
| Iron | Construction material (pure or alloyed) Energy carrier/storage | [37, 38, 41] |
| Magnesium | Construction material (alloyed) Solid powder as rocket fuel Energy carrier/storage | [33, 37, 38, 41] |
| Titanium | Construction material (pure or alloyed) Energy carrier/storage | [37, 38, 41] |
| Manganese | Construction material (alloyed) Energy carrier/storage | [37, 38, 41] |
| Chromium | Construction material (alloyed) | [35, 37] |
| Sodium | Thermal fluid/coolant Energy carrier/storage | [41, 44] |
| Potassium | Thermal fluid/coolant Energy carrier/storage | [41, 45] |

## Properties of Earth versus the Moon and relevant effects on processing

Current comminution, beneficiation, metal production, and standard business practices for mining on Earth have all been developed and refined over time with base assumptions that come inherently from the physical and chemical conditions found on Earth. With the intent to expand such operations to the Moon, these practices need to be examined in detail in terms of the likely effect(s) that these changed conditions will have on them. Here we will analyse the main metrics that differ between the Earth and the Moon, the immediate effects these differences will have, and some of the more esoteric follow on effects that need to be considered in order to establish industrial scale mineral processing and metal extraction processes on the Moon. A comparison of some of the key metrics is provided in Table 5. This table considers inherent natural factors such as gravity, pressure, average surface temperatures, and other key aspects. Human-induced factors involving supply chain costs and access to a human workforce, also warrant significant consideration and are discussed below.

*Table 5 - Comparison of the environmental effects of the Earth versus Moon on processing, adapted from Rasera [46], and Gibson and Knudsen [47]*

| Factor | Earth | Moon | Impact |
| --- | --- | --- | --- |
| Gravity | 9.8 m/s$^2$ | 1.62 m/s$^2$ | Impacts density separations; Reduces gravity-driven fluid/particle flow; Increases influence of non-gravity motive forces (surface tension, magnetic attraction etc.); Decreases head pressure, increasing potential pumping heights; |
| Average surface temperature | 14 °C (Day and Night) | 123 °C (Day) -178 °C (Night) | Temperature variation will affect particle properties e.g. Equipment must be able to survive extreme thermal variations. |
| Pressure | 1 atm. | Unmeasured (Day) 3x10$^{-15}$ atm (Night) | Fluids must be used in a closed system with artificial pressure; Reduction in energy requirements for dissociation; Convective cooling not an option outside of closed systems. |
| Human access | Abundant | Severely restricted | Highly reliable equipment (uncommon in terrestrial mining and metallurgy) is necessary. Modular designs will facilitate repairs/replacements. Automated robotic/remote controlled uncrewed equipment preferable. |
| Day/Night cycle | 24 hours | 708.7 hours | Regolith becomes less charged at night (no UV charging) Equipment cooling/heating is power intensive. Power generation at night becomes challenging, requires large power storage for solar based power solutions. |
| Water Availability | Plentiful | Rare Resource | Geochemical effects on feed material. Comminution and beneficiation processes need to be designed without the use of water. Renders the majority of terrestrial technologies useless. |
| Dust | Easily supressed | Electrostatic, abrasive, and everywhere | Dust suppression and/or mitigation technologies will become essential. |

*Gravity*

One of the most used and overlooked environmental factors in mineral processing and metal production technologies is gravity. The Moon has one sixth of the gravity of Earth at its surface, which has wide implications for most processes used in a mineral processing and metal extraction flowsheet.. Whilst some equipment in a generic mineral processing plant, such as slurry pumps, benefit from a lack of gravity due to the reduction of head pressures, the majority of the major processes utilise gravity as a motive force within the equipment or at least as a feed and discharge mechanism. Ball mills for example are unable to operate in microgravity environments, in reduced gravity environments they are theoretically usable but require significant upscaling. A ball mill of 10 m diameter would need to be increased to a diameter of 60 m for an equivalent power draw on the Moon [48], this is a size increase of 6 times and represents a large increase in required construction materials. Some equipment such as gravity separators and flotation cells have variants that use centripetal force, replacing gravity with the apparent centrifugal force, these equipment variants may be usable in reduced gravity environments with minor alterations to existing designs.

One potentially important side effect of micro-gravity conditions is the effect of surface tension on the wetting capabilities of ores and reagents. When considering heap leaching as a potential concentration technique, very rarely is the question raised of wetting capabilities, gravity and some minor agitation are usually enough to saturate all mineral surfaces when exposed to, for example, an acid. This assumption is challenged under microgravity and to a lesser extent under the reduced gravity conditions on the Moon. In reduced gravity conditions the surface tension of the liquid plays a larger role in the mixing process. Whilst 1/6 g may be enough to render this a non-issue on the Moon, the topic warrants at least cursory inquiry when considering the formation of a slurry.

It is possible to artificially induce gravity using apparent centrifugal force. Terrestrially, this method is only used to create increased apparent gravity environments which are used to increase separation efficiencies of processes predicated on separation due to density [1, 2]. The design of equipment that uses artificial gravity to simulate Earth like conditions (9.8 m/s$^2$ acceleration), whilst possible, is costly in terms of construction material mass and energy requirements for operation, and should thus be avoided if possible.

*Pressure*

When considering pressure on the lunar surface it is important to distinguish between absolute measured pressure and the expected pressure range of operation, as this will be increased by a prolonged human and robotic presence on the lunar surface. The absolute pressure measured at night on the lunar surface by equipment placed on the Apollo 14 and 15 missions was $3\times10^{-15}$ atm [49]. Daytime measurements were 'contaminated' by off-gassing from the equipment itself [49]. This leads to the clarification that whilst the number of $3\times10^{-15}$ atm given as the pressure for the natural lunar atmosphere at night is correct in theory, in practice, this pressure is expected to rise during the day [50].

Human or robotic activity will have a noticeable effect on the natural lunar pressure, however, any artificial increase in pressure will naturally dissipate over time [50]. Whilst not relevant for near term activity, when much larger industrial operations are considered on the Moon some thought will need to be given to gas discharge rates in terms of their effect on the atmosphere. It was estimated by Vondrak [50] that a constant artificial gas discharge rate (from venting, sublimation of volatiles during excavation/mining, rocket powered ascent and decent, and other factors) averaging above 10 to 100 kg/s would theoretically cause a transition on the Moon to a longer lived atmosphere which equates to pressures above $\sim10^{-10}$ atm [50]. Release rates of this magnitude will not be reached by small scale operations and infrequent rocket powered ascent/descents. For reference, each Apollo mission resulted in the release of $\sim10^4$ kg of rocket exhaust [51] which almost doubled the mass of the lunar atmosphere for a short period. A benefit to the natural loss rates is that if the artificial gas release is stopped (like in the case of a single rocket launch), the lunar atmosphere will naturally revert back to its original pressure over time [52].

The pressure on the Moon, or the lack of gas present under ambient lunar conditions, has varied physical and chemical effects. Take for example the standard passive cooling techniques used for most industrial pumps and motors which rely on convective heat transport to maintain safe operating temperatures. Under vacuum conditions convective heat transport is not a reliable method of cooling. Instead, equipment that includes conductive and radiative heat management will need to be implemented. Similarly, with a lack of ambient atmosphere, dust management becomes significantly harder. The effect of air resistance on the travel distance of falling dust and larger particles is easy to overlook. Agosto [53] whilst testing magnetic and electrostatic separation of Apollo 10 regolith (Sample ID #10084,853), concluded that for electrostatic separators, "Existing designs with minor modifications would probably work very well in a gas[pressurised] environment established on the Moon, but major modifications are required for efficient vacuum operation." [53].

There are also significant physiochemical effects. One of these is that of the effect of pressure on phase transitions between condensed and dispersed phases, namely that material will start to evaporate and sublimate at lower temperatures than required on Earth. Solids, as they are heated up, will start to directly sublimate into a vapour phase at these low pressures. Along with the effects on water-based processing mentioned in Table 5, this also affects molten phases. Whilst the total pressure does not affect the vapor pressure of, for example, oxides, and thus the kinetics of sublimation are very slow [54]; the lower temperature vaporisation of materials is thermodynamically favoured. This is one of the significant disadvantages to processes such as molten oxide electrolysis in which a molten phase is required as an electrical energy carrier, the operation of such a processes in the natural vacuum on the Moon will result in a steady, if small, loss of material to evaporation. Similarly, processes that require gas-solid reactions (for example hydrogen and carbothermal reduction) will require pressurised atmospheres to operate, as in these processes, pressure correlates to the rate of the reaction [55]. It is interesting to note that in such cases, unless un-protected human access is required, there may be benefits to creating an artificial atmosphere that is significantly higher in pressure than 1 atm to further promote the reaction kinetics.

The second chemical effect of the high vacuum environment of interest for metal production is its influence on metal compound stability. There is a direct correlation between total pressure and the energy required for metal compound reduction, namely less energy is required to reduce a metal compound into gaseous products at lower pressures [56]. An Ellingham diagram generated at the minimum measured pressure on the Moon can be seen in Figure 4. This diagram has been generated using the FactSage thermochemical modelling software package, more detailed descriptions of this diagram can be found in other work [56]. Further information on the modelling package can likewise be found elsewhere [57-59].

The Ellingham diagram in Figure 4 shows the low temperatures theoretically required for oxide reduction on the Moon. At equilibrium, all plotted oxides in an isolated system are in a reduced vapor phase above 1650 °C. Whilst the maintenance of any system at a pressure as low as $3x10^{-15}$ atm with gas evolution taking place is functionally impossible, this figure is and extreme example showing the theoretical effect of the ambient lunar pressure.

The field of vacuum metallurgy has been well studied historically and the use of vacuum in industrial processes has been successfully demonstrated [60]. These vacuum metallurgy processes tend to operate at 'low vacuum' pressures down to ~$10^{-5}$ atm, this lower pressure limit being due to the available high flow pumping equipment for industrial sized operations. On Earth, the energy required to pull a vacuum in a reactor is higher than the amount saved by this vacuums' effect on the reduction energy requirement [61]. Therefore, vacuum is used on Earth as a form of extreme inert atmosphere, i.e. to remove the complication of the gas inside the reactor reacting with one of the products or complicating the chemical reduction pathway [60]. The ambient pressure conditions on the Moon theoretically remove the requirement for, and limitation of, the pumping equipment; high vacuum conditions (<$10^{-10}$ atm) are readily available with minimal energy requirements due to pumping if an open reactor design is implemented. The ability to use this natural vacuum environment to reduce the energy requirements for metal reduction processes is relatively unexplored, but has great potential for future astrometallurgical applications [62].

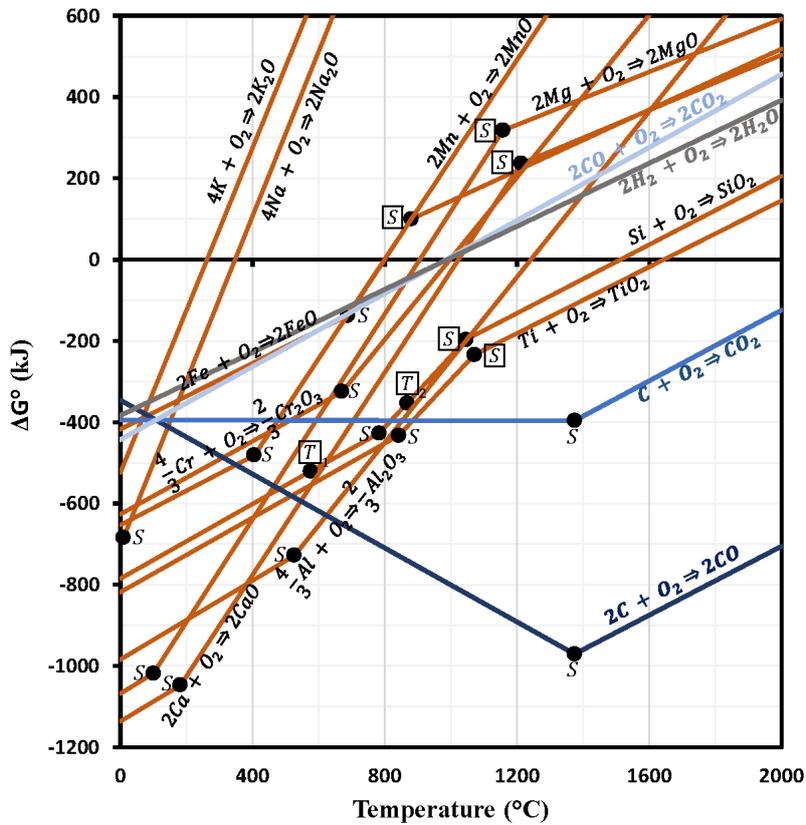

Figure 4 - Ellingham diagram plotted at a total pressure of $3 \times 10^{-15}$ atm. Diagram generated using FactSage 7.2 software package and relevant databases (FTOxid, FactPS).

*Day/Night cycle and Energy Generation*

The synodic period on the Moon, a lunar day, is 29.53 (Earth) days or 708.72 hours [63]. This results in two Earth weeks of constant access to sunlight and an equal time with no access to sunlight. The illumination on the lunar surface during the day averages 1361 W/m$^2$ [64] which varies throughout the year and decreases based on the angle of incidence. This is the same flux that hits the upper atmosphere of Earth, however, on the Moon this solar flux is not attenuated by the atmosphere and weather phenomenon.

Such a long duration for days and nights will cause significant issues with any continuous processing operation. Any solar powered processes, be that electrically using photovoltaic cells or directly using concentrated solar thermal energy, will have to be run in some variant of a long batch process. For a more traditional process setup (without a solar furnace) there exists the potential to run off nuclear energy or battery power during the night, however, the size of the batteries required may necessitate in-situ fabrication. A secondary option is the use of orbital solar power stations, also called solar power satellites (SPS) [39, 65], these are a potential energy solution in the future that operate by collecting energy in orbit and transferring it wirelessly down to the planet or moon. SPS may end up being an ideal technology but are yet to be implemented and require significant investment to set up without extra-terrestrially derived construction materials.

An alternative to nuclear power and SPS is to operate in the polar regions. Due to the Moons low inclination to the ecliptic, there exist areas on the lunar poles, specifically crater ridgelines, that have access to sunlight for the majority of the year. These areas, colloquially called the 'peaks of eternal light'[66], are illuminated for up to 82.9% of the year at ground level [67]. This increased access to sunlight, and subsequent ability to operate primarily from solar power sources has prompted significant interest in these geographical locations [67-71].

Regardless of the power source chosen, energy will be a significant resource that needs to be monitored carefully, especially when considering the large energy requirements usually associated with mineral processing and metal extraction activities [1, 2]. A detailed analysis for any proposed process or equipment for use on the Moon will need to be completed that considers the trade-off between production rate, energy requirements, and the efficiency of that process or equipment. The minimisation of electrical and thermal energy requirements within processes will be significantly favoured in industrial scale metal production facilities.

*Surface Temperatures*

The large amount of solar flux and the lack of air for convective heat transport result is a large natural temperature range on the Moon. The temperature in equatorial regions ranges from 123 °C during the day to -178 °C at night [72]. The high temperature is reached when a material reaches thermal equilibrium, this low temperature however does not represent thermal equilibrium and is instead reached just before sunrise [72]. This is due to the fact that when shadowed, heat will continue to radiate away from a substance. In PSRs where thermal equilibrium can be reached, this results in cryogenic temperatures of <-220 °C [72]. Whilst not to this extreme, there is a notable temperature drop between illuminated and shadowed areas even during the day. These extremes in temperature can cause a significant issue for industrial activities. This high temperature range between illuminated and shadowed areas will necessitate careful design and material selection of any industrial equipment meant for use on the surface.

During the day, when illuminated parts of equipment will experience high temperatures, and shadowed parts, low temperatures. The ability to regulate temperature will be important. Equipment will need to either be able to operate with large temperature gradients within the structure or utilise thermally conductive substances, like heat pipes and coolants, connected to radiators in order to manage heat build-up. One method that was used on the Apollo missions during transit between Earth and the Moon to minimise thermal loading on the craft was an intentional slow roll. This slow roll of the craft termed Passive Thermal Control (PTC), balanced the thermal loading due to sunlight and stopped any one side of the craft from getting too hot [73]. Where equipment cannot be designed to operate with large temperature differences within its structure, PTC-type mechanisms may prove useful.

Temperature variation between night and day poses an entirely different issue on extraction equipment. Any equipment that will be operated outside of an artificial atmosphere will need to be able to withstand extended periods of cryogenic temperatures during the night. Such temperatures are known to significantly decrease the strength of some construction materials, such as high strength steel [74]. The thermal control of the core essential electronics on lunar rovers has been accomplished historically by the use of a 'warm electronics box' [75]. This is a system that during the day will shed excess heat from the system to a radiator. At night, the connection to the radiator is cut and heat loss is minimised, extra heat is provided by heaters to maintain the required operating temperature [75]. This method works well but is currently only implemented for the small units attached to rovers. For industrial sized mining equipment significant modifications will need to be made. This required heating of areas that contain temperature sensitive material, will represent an increased electrical load during the night. The need to include temperature regulation systems in all equipment is critical for operation but should not present significant issues in terms of processing options.

*Water Availability*

A primary concern for the conversion of terrestrial mineral processing technologies for use in space and specifically on the Moon is the significant lack of available water. With the exception of some mineral sands processing techniques (e.g. electrostatic and magnetic separation), the vast majority of mineral processing technologies rely heavily on water for operation. Whilst water is available on the Moon in PSRs [24-26, 76], this water will be an expensive commodity in its own right [28]. Assuming a source of water can be found to enable the use of water in the metallurgical processes, a second significant issue arises when considering the physical effects of the vacuum on any liquid substances as mentioned above. The pressures present on the lunar surface lie well below the triple point of water [77], this means that in order to stop the spontaneous evaporation of the water the pressure of the processing system would need to be artificially increased. Similarly, in order to stop solidification or vaporisation of the water at average day and night temperatures on the Moon, the system would need to be heavily temperature regulated. Given the predicted costs of water on the Moon [28], and the difficulty of creating large, pressurised, temperature regulated environments in which to use it, it is important to consider alternatives to water-based mineral processing technologies.

*Dust*

The Lunar dust is the fine (<50μm) portion of the regolith material [8]. The lunar dust is very abrasive and, due to its glassy composition, fine size, and large surface area to volume ratio, is extremely prone to building up an electrostatic charge [78]. Due to its propensity to build up a static charge, and the interaction between this charge and the natural plasma sheath found near the lunar surface, the dust has been found to naturally levitate [79]. This levitation is affected strongly by the day night cycle and as such, the terminus between day and night is accompanied by significant dust movement [78]. The natural levitation of the dust, along with its electrostatic charge make is impossible to operate on the lunar surface without the dust adhering to most open surfaces causing multiple issues. A NASA report on the effect of the dust identified nine main areas in which the dust was, on the Apollo missions, and would continue to be, for future endeavours, an issue in terms of its effect on Extra-Vehicular Activity (EVA) on the Moon [80]. These nine areas were:

- Vision obscuration,
- False instrument readings,
- Dust coating and contamination,
- Loss of traction,
- Clogging of mechanisms,
- Abrasion,
- Thermal control problems,
- Seal failures, and
- Inhalation and irritation.

It was noted by Gaier [80] that "The severity of the dust problems [on the Apollo missions] were consistently underestimated by ground tests…". Dust will be a significant issue for any operation on the Moon, industrial or otherwise. Whilst the presence of dust should not invalidate any specific processing technologies, the general awareness of its effects and modification to equipment to enable a minimisation of its impact will be critical. Of specific concern is the protection of any rotating or mechanical equipment. Due to the highly abrasive nature of the dust, the protection of joints and rotating surfaces, prevalent in mineral processing operations, will be a complex problem. The dust also poses significant potential issues in terms of human health similar to those associated with silicosis, there is concern over the lack of understanding as to the toxicity of the lunar dust on humans [81]. Technologies that can be used to shield or remove dust from equipment [82-84] will be extremely beneficial for both crewed and un-crewed lunar industrial activity.

*Supply Chain Issues*

One of the primary arguments for SRU is Supply chain lead times and the reduction in the mass of the payloads carried up to orbit from Earth, often referred to as the up-mass. Up-mass minimisation is critical to the economic viability of any mission due to the related transport costs. Current estimates of USD $35,000 per kilogram from Earth to the lunar surface (bulk haulage) [28, 85] are expected to significantly reduce in coming years due to the privatisation of launch capacities [86, 87]. However, the economic argument for ISRU relies on the fact that the cost of transport from Earth is extremely high as compared to the cost of production of the same resources from local materials. This renders the idea of transporting equipment for terrestrial type mining and processing up to the Moon almost impossible from an economic standpoint [88]. Regardless of cost, the transit time for a cargo shipment using current technologies is large. The Apollo missions took in excess of 100 hours from launch to touch down on the Moon [73]. Fuel optimised trajectories, which cost less in fuel, can take months; SpaceIL's robotic Beresheet mission took a total of 48 days before hard landing on the lunar surface [89]. Such long lead times render emergency re-supply a non-option for most operations and play heavily into some of the health and safety concerns that will be explored below.

The cost of transport to the lunar surface is often cited as a justification for ISRU but simultaneously often forgotten in the proposal of processes for use in this field. This is especially notable in terms of extraction and reduction processes that use chemical reagents. In terrestrial processes the use of reagents is preferable if they generate a significant gain in product grade or recovery. However, supply chain logistics will become a significant issue in the operation of an extra-terrestrial processing facility. It is relatively simple to take a well-known terrestrial process, apply a lunar feedstock composition, and decide that this process is possible on the Moon. Ignoring the complexity of process modification to account for gravity as explored above, this paradigm ignores the significant cost involved with on-going reagent use. Some processes, such as hydrogen reduction of ilmenite, and carbothermal reduction, can theoretically be designed to recycle most of the reagent (H and C respectively), others such as solid electrolysis use an electrolyte as a reagent that can also theoretically be recycled with quite good recoveries. All of these processes will however need re-supply at industrial scales. Even small losses at laboratory scales, less than a percent for example, will result in tonnes of reagent loss after scale-up and will represent millions of dollars in transport costs if a terrestrial reagent source is used for resupply. Herein lies a challenge in the repurposing of terrestrial processing technology for extra-terrestrial use. These processes, though very well-known and at high technology readiness levels (TRL), are not optimised for use in space. Other processes, such as thermal dissociation and molten regolith electrolysis, that do not use reagents, are not well understood and require significant research in order to raise their TRL.

Due to the launch costs involved in sending the material and equipment required to build a processing facility on the Moon, i.e. the up-mass cost portion of the capital expenditure (CAPEX), it is reasonable to conclude that it is economically beneficial for processing facilities to designed to be lightweight in nature. This compliments the assertion that a simple process schematic is best. A simple process schematic is good from an automation perspective, but also in terms of the required up-mass to begin processing operations. These considerations will of course change drastically when a lunar originating metallic feed stock is available for process plant construction/expansion.

In terrestrial industry, the ready access to wear parts means that regular scheduled maintenance and replacement of such parts is considered an adequate measure to save on costs due to equipment failure and downtime. Unless wear parts can be manufactured locally, the cost of transporting these parts from Earth is prohibitive. Equipment design favouring long lasting operation will quickly become far superior to easily replaceable parts. This is particularly important when considering comminution circuit design which historically leans heavily on replaceable wear parts [1].

It is important to note that whilst most of the 'detrimental' effects of processing technologies mentioned herein, especially in regards to temperature and pressure, could be attenuated simply by creating a closed system in which such temperature and pressure are carefully controlled, the inclusion of the equipment required for a closed system would represent significantly higher equipment mass and thus initial payload mass and CAPEX estimates. Since the premise of ISRU technologies is specifically to *reduce* the required payload mass for extended habitation and exploration activities in space, this solution seems counterproductive. The concept of minimising launch costs will be referenced many times in this analysis and until equipment and resources can be routinely sourced from industries based off Earth, the cost of material transport from Earth to the Moon will severely limit the economic viability of many processing options.

*Radiation*

The surface of the Moon, somewhat protected from solar radiation by the Earths' magnetosphere, experiences considerably higher radiation levels than the surface of Earth. Radiation on the lunar surface consists of electromagnetic (EM) radiation and energetic particles from the sun, as well as galactic cosmic rays [90]. On top of the background solar radiation, periodically large bursts of particles from the sun, termed solar particle events, can also impact the lunar surface. These solar particle events can be many orders of magnitude larger than the background radiation, and are very hard to predict [91]. On Earth, cosmic radiation accounts for an average of 0.4 mSv over a year in a human. On the Moon this number is estimated to increase to between 110 and 380 mSv per year depending on the solar cycle, and a single solar particle event can result in a potentially lethal dose of up to 1 Sv [91].

This radiation has significant repercussions for both human habitation and material properties over time. The effect on human safety will be considered in the next section, however, both UV radiation and ionising radiation (protons and electrons) can significantly degrade construction materials, especially in the case of polymers [92, 93]. Ionising radiation can also damage electronic components and solar cells [92]. Though radiation does not impact lunar feedstocks the development of future processing equipment, and the requirement for shielding of some components from radiation will be essential.

*Human Access and Health and Safety*

Safety considerations are always a primary concern in the design and operation of terrestrial processing facilities, this will be no different in space. In fact, with a myriad more risks to consider for any operation, these risk assessments, both for occupational health and safety (OHS) and to mitigate the potential of equipment damage will be significantly more complicated. The use of human operators is assumed for terrestrial processing operations, and in space, requires a liveable habitat and all the health risks inherent in living in a pressure vessel in low gravity. The risk posed by radiation exposure to human crew is severe and requires significant shielding of a habitat to avoid lethal doses [91]. Similarly, the risk of a meteoric impact event, whilst a low likelihood, could have catastrophic effects in terms of OHS and equipment damage. One suggestion regarding mitigating such risk is to build habitats and facilities under the lunar surface itself [94-96]. Such an operation might entail more complicated energy arrangements, and communication infrastructure, but would protect equipment and humans from exposure to meteorites as well as radiation.

Operations designed for human access would presumably be run in a manner similar to current Fly In Fly Out (FIFO) operations on Earth. However, these operations can be extremely large due to the myriad roles required to keep a camp operational. The most obvious answer is to move to automated machinery with the majority of non-automatable tasks able to be completed remotely from Earth, or in lunar orbit, with only troubleshooting and unforeseen maintenance completed by in-situ workers or rapid landing teams from an orbiting station.

*Remote Operation/Automation*

Due to the issues with human access, and the cost of transportation, the ability to remotely operate processing equipment and fully automating as much of the process as possible will be essential for extra-terrestrial processing operations. An ideal space-based industrial processing facility would be designed to operate without on-site human operators. Such a facility, while simple in theory, has never been done on Earth to date at any significant scale. Underground mining automation has slowly been progressing with safety and productivity as a motivating factor, and has seen great success in the cases of BHP and Rio Tinto operations in Australia [97, 98]. Mineral processing operations, however, are currently inherently too complicated and fickle to fully automate. Some success has been seen with the development of digital twins for prediction purposes [99] but this technology is a long way from allowing for the full automation of a minerals processing plant. It is unlikely that full automation will be possible, especially considering the certain eventuality of unforeseen maintenance requirements. It is interesting to consider however the pursuit of this type of processing facility and the alterations to plant design it would necessitate.

Automation, whilst considered useful on Earth, comes secondary to production and cost. Human labour is comparatively extremely cheap on Earth and most risks considered manageable, however with expansion into space, remote operation will become more economical. The idea of designing a processing plant with complete automation (including maintenance and troubleshooting), as a pre-requisite is a challenging prospect. This goal is a reasonable target when considering the aforementioned risks associated with human presence. An automated or even remote-controlled operation can be designed without all the requirements for human habitation and subsequent infrastructure, launch mass, and increased CAPEX requirements. This would represent a significant increase in economic viability of an SRU operation. In the pursuit of automation, it stands to reason that the most optimal processing plant will be a schematically simple process. The fewer moving parts, the fewer complicated interactions, the fewer pieces of critical equipment required, the more likely it is that a plant could, in theory, be automated.

*Recycling and waste generation*

Waste, or gangue, generation and handling is a significant part of any mineral processing operation. Terrestrially, waste materials from processing operations are often stored in large tailings dams that allow for water reclamation [100]. On the Moon, where materials handling needs to be minimised wherever possible to save on costs, such tailing materials should, where possible, be treated as a target for recycling or re-use. As presented by Naser [37, 38], the most obvious potential use for metal-depleted 'gangue' material is as a feedstock for additive manufacturing via sintering, or as a component in waterless concretes, for construction purposes. The reuse of previously handled materials in processing, and the recycling of industrial and human waste also fall within the expansive umbrella of ISRU [101]. In the context of the current work, we can conclude that processes that minimise waste or produce gangue streams in a form useable in other extra-terrestrial construction endeavours are preferable. Similarly, metal extraction processes that can operate with modified feed compositions that include industrial waste streams will be extremely valuable in promoting a more cyclical economy on the Moon.

# Comminution

The initial step of most mineral processing operations is the comminution of the feed material [102]. In terms of equipment and processing plant design, the primary issues with the conversion of terrestrial comminution circuits for use on the Moon are: water availability, dust generation, energy requirements, supply chain costs (for wear parts), and gravity. Of these, gravity is the primary concern as it is used as a feed and discharge mechanism for all dry equipment and is the primary driving force of a number of processes. Table 6 shows some common comminution equipment and compares the effect of gravity on Earth to that found on the Moon. A 'usability assessment' for each piece of equipment has also been supplied.

*Table 6 - The effect of gravity on generic comminution equipment*

| Equipment | Use of gravity | Effects of reduced gravity | Usability assessment |
|---|---|---|---|
| Pressure Crushers (Jaw, Gyratory) | Feed and discharge, keeps rock from discharging upwards instead of crushing, dust suppression | Crushing efficiency and throughput negatively affected, dust may travel a long way or be suspended for longer periods of time | Medium (requires re-design) |
| Impact Crushers (Vertical Shaft Impactors, Hammer Mills) | Feed and discharge, keeps rock from discharging upwards instead of crushing, dust suppression | Throughput lowered due to feed and discharge rates. Crushing mechanism not affected. Increased dust generation. | Good (requires re-design but main mechanism not affected) |
| Conveyor | Promotes friction and thus mobility on belt | Material may be more prone to rolling, less power required for belt operation | Good (may be more efficient than on Earth, may require design adjustments to stop material rolling) |
| Vibrating Screen | Feed and discharge, Used as driving force | Significantly lower screen efficiency, higher likelihood of dust evolution | Poor (requires significant alterations to current designs) |
| Mill (Rod, Ball, SAG) | Feed and discharge, primary driving force | Mills would need to be designed significantly larger for similar efficiency and throughput | V. Poor (requires alternate technology) |
| Hydrocyclone | Aids separation efficiency | Orientation of cyclone less important, will require alterations to operating pressure/flowrate | Good (minor adjustments may be beneficial) |

Less common in industry, crushing and grinding methods such as vertical shaft impactors, that use centrifugal force as the main driving mechanism may be usable; however, even these impact crushers use gravity as a feed mechanism and require the regular replacement of wear parts. Similarly the use of electric pulse fragmentation [103] may find use in extra-terrestrial comminution processes, gravity would not affect this equipment beyond feed and discharge however this technology is itself at a low TRL, requiring further development.

Dust generation and supply chain costs are critically important but can be accounted for with significant planning. Like gravity, the lack of water availability is a significant hurdle for the use of classical comminution flowsheets on the Moon. The lack of available water renders wet comminution techniques, (i.e., grinding processes using a ball mill and hydrocyclones for classification) unpractical. This requires the development of non-standard grinding techniques if traditional particle sizes are still required for the beneficiation and reduction of the lunar ores. When considering the use of traditional comminution methods, it is also important to consider the power requirements usually associated with the comminution circuit. On average ~29% of total electrical energy consumption for modern mining and processing operations is due to the comminution circuit alone [104]. The minimisation of electrical energy use is beneficial on the Moon due to the mass and subsequent launch cost associated with large power generators. Economic operation in space, and specifically on the Moon, will significantly favour the use of more energy efficient equipment.

In terrestrial operations the primary concern of the comminution circuit is that of particle liberation [102]. When considering the processing of lunar regolith material, that has a naturally occurring P80 of 257µm (see Figure 2), most large mineral grains present in the lunar regolith are already liberated to some degree [16]. The main contender for non-liberated particles is that of the prevalent glasses and agglutinates formed by meteor impacts. The difficulty in deciding how to processes these particles is that the glass composition very closely matches that of the bulk regolith [16], and there are no grains within that can be liberated further. With a glass and agglutinate composition of between 10 and 60 percent [16], the instant rejection of this portion of the regolith due to its glassy and/or agglutinated form will result in a targeted rejection of 10 to 60 percent of the targeted elements. This will not directly affect a mineral concentrate grade but will significantly affect target metal recoveries when calculated from bulk feed composition. The inability to liberate minerals from the glassy fraction of the regolith, and the existing particle size of the natural regolith call into question the necessity of operating a comminution circuit at all.

An argument worth consideration is that of eliminating the comminution circuit from lunar operations entirely. The average particle size of the lunar regolith is already akin to that attained by standard comminution processes, with a large proportion falling into the category of 'fines' (<20 µm) that is historically a hinderance to standard mineral processing operations [102]. Similarly, comminution circuits are often the cause of significant dust generation, which in a low gravity environment and with the high susceptibility of the regolith to static charging will result in adverse operating conditions. Using classification equipment such as the vibrating classifier proposed by Kawamoto [105], an electrostatic travelling wave as suggested by Adachi *et al.* [106], or a centrifugal sieve as demonstrated by Wilkinson [107], it may be beneficial to remove large particles/rocks, selectively sinter the fines into usable sizes, and develop beneficiation and metal reduction processes that can use slightly larger particle sizes as a feed material. The elimination of the bulk of the comminution circuit in a mineral processing operation would significantly reduce the energy requirements of the process, eliminate a number of wear parts that would need regular replacement/maintenance with current technologies, and simplify the plant schematic increasing economic viability and the potential for plant automation. The development of usable comminution equipment is still required for mining operations targeting bedrock and any potential mineralised areas found within, but for initial processing endeavours targeting regolith material as a feedstock, the complete elimination of a classical comminution circuit and replacement with basic particle size classification has enough incentives that it is worth detailed investigation and cost benefit analysis.

# Beneficiation

The second major step in a mineral processing operation is the beneficiation of the (usually) comminuted material. Beneficiation processes result in one or more concentrate streams, and a gangue or waste material stream [102]. The primary concerns regarding the conversion of terrestrial beneficiation processes for use on the lunar surface are water availability, supply chain costs, gravity, and pressure. The effect of these altered conditions on traditional beneficiation equipment can be seen in Table 7, a usability assessment has again been given to each equipment type.

Static gravity concentration or separation techniques (classic jigs, tabling, etc.) suffer heavily in lunar conditions as they rely on gravity as the main separation mechanism. Equation 1 shows the acceleration (dx/dt) of a mineral particle with density $\rho_s$ in a fluid with density $\rho_f$, in a gravitational field (g) [108]. Since separation efficiency in static gravity concentration techniques is directly related to the difference in acceleration between particles, a lowering of the gravitational field (g) will significantly reduce this separation efficiency.

$$\frac{dx}{dt} = \left(\frac{\rho_s - \rho_f}{\rho_s}\right) g \qquad 1$$

The beneficiation processes that are more strongly affected by this reduction in gravity tend to have centrifugal variants that could be used instead. Centrifugal variants of equipment for flotation [109], dense media separation [110], and gravity separation [111] have all been studied in depth for terrestrial use. Equipment such as gravity spirals lie in a grey zone as they use gravity as an essential motive mechanism, but the actual separation achieved is due to the generated centrifugal force. Secondary equipment such as slurry pumps and storage tanks actually benefit from the lower gravity as less energy is required for overcoming head pressures and particle settling respectively. For beneficiation, the altered gravity conditions do not seem to pose an insurmountable issue. Of significantly greater concern is the heavy reliance of most traditional beneficiation processes on the access to and use of water.

Water is used in the majority of terrestrial mineral processing plants from the grinding stage of comminution onwards [1, 2]. On the lunar surface the use of water, and hydrometallurgical processes in general, has several issues. The significant lack of water available and the cost of said water in cislunar space is prohibitive. Similarly, these hydrometallurgical processes often utilise chemical reagents (acids, modifiers, collectors, dense media, frothers, dispersants, flocculants, etc.) which for ongoing operations represent significant re-supply costs. An added complication explored above is the requirement for artificial pressurisation for any processes including liquids. Due to these combined factors, the use of hydrometallurgical processes on the Moon is likely to be uneconomical. This leaves only the options of dry beneficiation, processes such as electrostatic separation, magnetic separation, particle sorting, and dry gravity separation techniques. These processes are often associated with mineral sands type processing facilities.

The need for dry beneficiation techniques has not been ignored by the ISRU community. Rasera *et al*. [112] in a recent thorough review of lunar beneficiation techniques covered solely dry methods of beneficiation. Significant research has gone into the dry beneficiation of lunar regolith, Rasera *et al*. [112] noted that most lunar specific research to date has been focussed on the concentration of ilmenite ($FeTiO_3$), and has ignored the <50μm size fraction of the material. The majority of this research has been into electrostatic separation [53, 106, 113-123], with some into magnetic separation [124-127] and size sorting [107]. Of these, seven test series have been conducted on returned lunar samples #64421 [125], #10058 [124], #71055 [124], #10084 [53, 120], #14163 [114], and #70051 [114]. With such a comprehensive study of ilmenite concentration, it will be beneficial for future research to consider the concentration of other prevalent minerals on the lunar surface such as anorthite ($CaAl_2Si_2O_8$).

*Table 7 -Traditional Beneficiation Equipment Considerations for Lunar Use*

| Equipment type | Effect of gravity | Lack of water | Ambient pressure | Supply chain (wear parts, reagents) | Usability assessment |
|---|---|---|---|---|---|
| Flotation | Will lower buoyancy force significantly lowering concentration efficiency | Cannot operate without water. | Uses water, requires artificial pressurisation. | Large amounts of reagent required (non-recoverable). | V.Poor (Primary separation mechanic severely impacted, requires water, large supply chain costs) |
| Centrifugal Flotation | Gravity replaced with Centrifugal force. Less pressure required to counteract natural gravity | | | | Poor (requires water, large supply chain costs) |
| Traditional Gravity (Tables, Jigs, etc.) | Separation efficiency significantly impacted | | Water fluidised and pneumatic variants need artificial pressurisation. | | V.Poor (Primary separation mechanic severely impacted, require water, requires alternate technology) |
| Centrifugal Gravity Separators | Gravity replaced with Centrifugal force. Material discharge gravity driven, so slower in low g | | | | Good (Requires re-design, wear parts a consideration) |
| Magnetic Separator | Material flow caused by gravity; separation not affected but re-design required. | Waterless operation variants | Particle trajectory alteration due to lack of air resistance. | Wear parts require regular replacement (abrasion) | V.Good (Requires re-design, wear parts a consideration) |
| Electrostatic Separator | | Does not require water. | | | V.Good (Requires re-design, wear parts a consideration) |
| Sorting Machines | | | | | Good (will require re-design, only applicable to course material) |
| Dense Media Separator | Will lower buoyancy force significantly lowering concentration efficiency | | Uses liquid media, requires artificial pressurisation. | Requires dense media reagent replenishment. | V. Poor (Gravity affects primary concentration, significant reagent replenishment required) |

Electrostatic and magnetic separation have significant promise for lunar beneficiation activities. For electrostatic beneficiation, the lack of ambient gas, in addition to the air resistance effects, will impact the charging properties of particles [53, 118]. The design of the charging mechanism for this process will need detailed scrutiny, this is in addition to including the effect of gravity on the process which, with the exception of Quinn *et al.* [113], has had very little study to date. Another potential issue for the use of electrostatic separation is the nature of the lunar environment to naturally statically charge the regolith [79]. Pre-charged regolith, whilst already being a massive issue in its own right [84, 128], could render separation techniques based on artificial charge acquisition useless; as such, de-charging of the bulk regolith prior to feed into an electrostatic separator would be essential.

Along with electrostatic and magnetic separation techniques, sorting machines pose an interesting beneficiation (or classification) prospect. One property of the regolith that significantly affects the potential use of sorting machines, but also other dry beneficiation techniques, is that of the particle size. Figure 5 shows the accepted feed particle sizes of 27 beneficiation techniques separated into wet (upper) and dry (lower) methods. Note that the minimum feed size for the dry processing techniques is 70 μm, with the exception of dry cyclones (50 μm), this is around the average P50 of the regolith material from the data set presented in Figure 2 (69 μm). This means that using these existing dry processing techniques, a classification stage that selectively removed 50% of the regolith material (the fines <70 μm) before beneficiation would be required. It is also interesting to note that some of these dry techniques still require fluidisation using air and as such would still need artificial pressurisation as covered in Table 7.

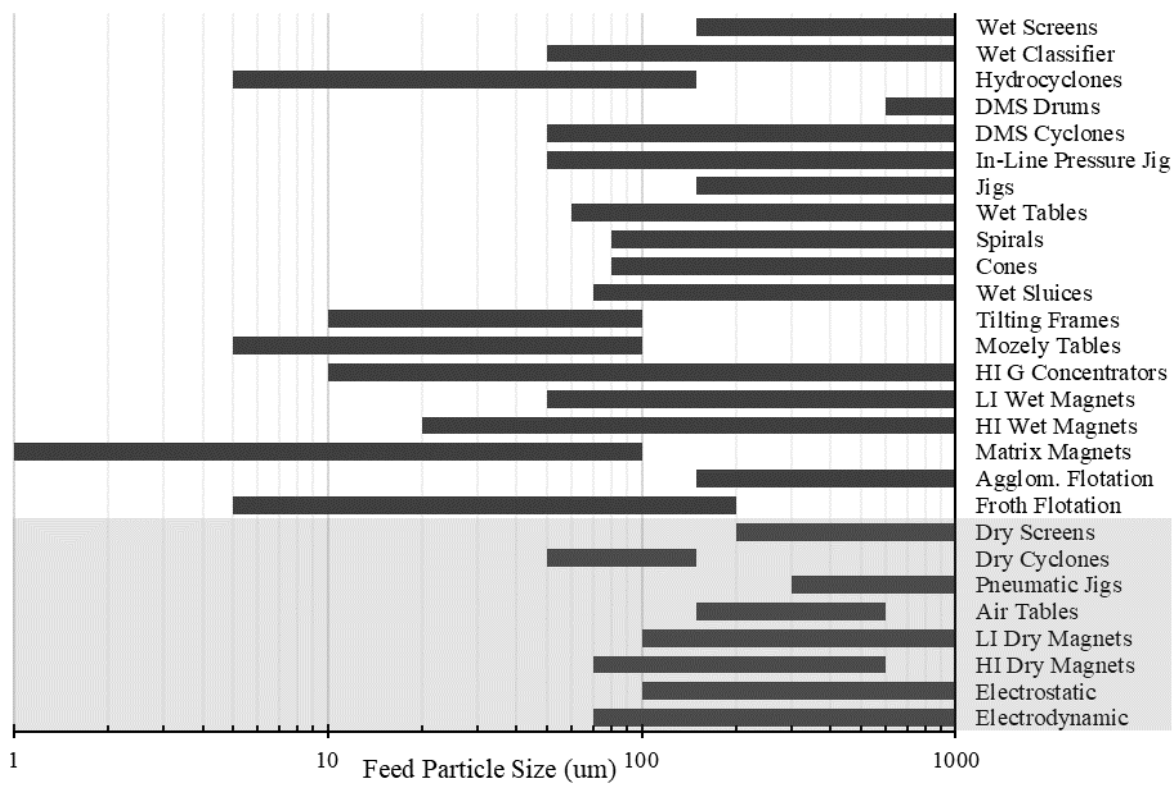

*Figure 5 – Wet (upper) vs. Dry (lower) processing options and their accepted feed particle sizes, reproduced from Nunna et al. [129]*

In addition to particle size, the existence of the lunar regolith in a much more modally homogenous state (i.e. with fewer liberated minerals) than most beneficiation feedstocks on Earth, significantly reduces the effectivity of a beneficiation process. Predicated on the separation of material based on the different physical and chemical characteristics of the target and gangue minerals, the large amount of glass present in the regolith, representing a mixed amorphous silicate-type material, render traditional beneficiation techniques unusable. Even newer proposed techniques such as electrophoresis for mineral separation [130, 131] suffer from the large percentage of amorphous material present. In a similar argument as was made for the comminution circuit; the elimination of the beneficiation stage of processing warrants detailed consideration. This would need to be the topic of an entire study, however it is not unreasonable to assume that the added process complexity inherent in a beneficiation process, the required up-mass, subsequent increase in CAPEX, increase in process complexity and thus difficulty in automation, and the inability of a beneficiation stage to consistently process the majority of the regolith either due to particle size or composition, result in the elimination of this circuit being a potentially economically incentivised endeavour. Such eliminations would require the modification of current metal reduction technologies to be able to deal with significantly lower target material feed grades. The development of a metal reduction technology that can accept un-beneficiated feed material at non-standard sizes certainly has some significant benefits over current mineral processing and reduction flow sheets.

## Metal extraction

Metal extraction, or the reduction of oxides, is the final step in the production of usable metal resources. Historically, the focus of the ISRU community has been on the production of oxygen rather than metals [5, 6, 30, 132, 133], however, the by-product of the production of oxygen from an oxide is usually a metal alloy. The current work will focus specifically on metal production rather than oxygen production. That said, the majority of ISRU related research referenced here has treated the metal produced as a by-product rather than a target resource. Despite the focus that will be given to metal extraction in the current work, the significant increase in economically viability of a process if the oxygen liberated in the oxide reduction is collected as a saleable/usable product cannot be understated.

The primary concerns in the evaluation of metal reduction processes for use on the Moon are reagent usage and supply chain costs, energy requirements, feedstock availability, and process complexity. The analysis presented in the current work aims to ignore the TRL of the processes mentioned herein, instead the potential of each process based on the above-mentioned concerns will be analysed. A detailed review of the historical work that has been conducted in this field is listed in Table 8. The table details each process and separates studies based on target resources, temperatures of operation, and in some cases reagent architectures. A generalised usability assessment has again been provided for each general process. The authors would like to note that the studies and subsequent conditions and reagent requirements referenced in Table 8 represent a compilation of all published research in this area and the quality of the processes presented in the cited literature. Some studies referenced are preliminary studies and/or lack significant details but are included in the interest of portraying all the historical work in this area.

The majority of work in this area has focused on the design of individual reactors rather than considering an end-to-end process as a whole. With this said, the testing of individual reactors and extraction techniques are important for inclusion of these extraction processes in a larger mining flowsheet. As can be seen in Table 8 many candidate processes for oxygen production have been proposed. Of these, due to their high TRL, carbothermal reduction, hydrogen reduction, and molten oxide electrolysis (MOE) have been identified as prime candidates for experimental testing on the Moon during the Artemis program [30, 134].

Three broad categories of reduction can be seen in Table 8. Reduction using chemical potential energy, reduction using electrical energy, and reduction using thermal energy. Each of these categories will be discussed in general as each have advantages and disadvantages. Detailed descriptions of each process can be found in the references provided in Table 8.

*Table 8 - Possible metal extraction techniques for lunar use to date*

| Process | Research Target Resource | Useful By-products | Refs | Condition Requirements | Reagents/Required wear parts | Overall assessment |
|---|---|---|---|---|---|---|
| Carbothermal Reduction | Oxygen<br>Oxygen<br><br>Oxygen | Fe<br>Fe, TiC, $Fe_xSi$, P<br>Fe, SiC/SiO | [135]<br>[136-139]<br><br>[140-145] | <1100 °C<br>>1200 °C<br><br>>1600 °C | Carbon (Solid, Methane, CO) | Medium process complexity, regents required, requires artificial atmosphere |
| Hydrogen Reduction of Ilmenite | Oxygen | Fe | [27, 47, 146-156] | 700-1100 °C | Hydrogen | Medium process complexity, regents required, requires artificial atmosphere, limited feedstock availability |
| Molten Oxide Electrolysis (MOE) | Oxygen<br><br>Oxygen, Al-Si Alloy | Al, Si, Fe, Ti Alloy | [142, 157-162]<br><br>[163] | 850C-1250 °C<br><br>960-980 °C | Electrolyte ($CaCl_2$ - CaO, or $SiO_2$ - $B_2O_3$ – $Na_2O$, LiF, $CaF_2$, $MgF_2$, $Na_2O$, $NaBO_4$, $Na_3PO_4$, $Na_5P_3O_{10}$, NaF – $AlF_3$)<br>Anodes | Medium process complexity, regents required, requires partial artificial atmosphere, high electrical load |
| Molten Regolith Electrolysis (MRE) | Oxygen | Fe, Si | [145, 158, 164-174] | 1300-2000 °C | Anodes | Medium process complexity, electrodes required, requires partial artificial atmosphere, high electrical load |
| Solid Electrolysis | Oxygen | Mixed Metal Alloy | [175, 176] | 900-950 °C | Electrolyte ($CaCl_2$ – CaO, $AlF_3$)<br>Anodes | Medium process complexity, regents required, requires partial artificial atmosphere, lower electrical load |
| Thermal Decomposition *<br><br><br><br><br>(& Selective Ionisation) | Oxygen<br>Entire Composition<br>Fe, Na, K<br>Oxygen<br>Oxygen<br>Oxygen, Si<br><br>Oxygen | Fe,<br>Oxygen<br><br><br><br>Metals | [177-187]<br>[188]<br>[62]<br>[189]<br>[190]<br>[155, 191]<br><br>[172, 185, 192, 193] | 1200-2000 °C<br>900-1800 °C<br>900-1200 °C<br>1427-1827 °C<br>2230 °C<br>2700 °C<br><br>>7000 °C | Hydrogen | Medium process complexity, no regents required, low electrical load (requires access to sunlight) |
| Ionic Liquid & Aqueous Electrolysis | Oxygen, Metals | | [194-196] | 0-300 °C | Ionic Liquids**,<br>Water, Electrodes, Sulfuric acid or Phosphoric acid in some cases. | High process complexity, regents required, requires artificial atmosphere, medium electrical load |
| Aluminothermic Reduction | Oxygen, Fe, Ti, Si, and Al, Mg<br>Oxygen, Si, Al, Ca | | [155, 197]<br><br>[198, 199] | 900-1000 °C<br><br>? | Conducted in electrolyte (NaF, AlF)<br>Anode ($Fe_{0.58}$-$Ni_{0.42}$)<br>Initial reactant (Al) | High process complexity, regents required, requires partial artificial atmosphere, high electrical load |
| Lithium Reduction | Oxygen | Mixed Metal Alloy | [200, 201] | 900 °C | Electrodes ($FeSi_2Li_x$, Pt, $La_{0.89}Sr_{0.1}MnO_3$), Reactant (LiF, LiCl or $Li_2O$), Electrolyte ($ZrO_2$), Catholyte ($La_{0.89}Sr_{0.1}MnO_3$) | High process complexity, regents required, requires partial artificial atmosphere, high electrical load |

| Acid Reduction | Oxygen | Fe | [142, 143, 145, 162, 202] | 20-110 °C | HF, $H_2SO_4$ | High process complexity, regents required, requires artificial atmosphere, limited feedstock availability**** |
| --- | --- | --- | --- | --- | --- | --- |
| Fluorine Reduction | Oxygen | Mixed Metal Alloy (Si, Fe, Ti, Al, etc..), CaO, MgO | [203-207] [172] | 500-750 °C 900 °C | KF, LiF, NaF, or HF | V. High process complexity, regents required, requires partial artificial atmosphere, high electrical load |
| Bio-Reduction | Fe(II) | N/A | [208-210] | 20-40 °C | Bacteria, Water, Defined Minimal Medium *** | High process complexity, regents required, requires artificial atmosphere, needs high precision temperature control |
| Carbochlorination | Oxygen, Al, Fe, | $TiO_2$, CaO, $SiO_2$ | [144, 172, 198] | 675-770 °C | Cl (g), Carbon (s), Hydrogen and /or water | High process complexity, regents required, requires partial artificial atmosphere, medium electrical load, specific feedstock required |
| Calcium Reduction | Oxygen | Metal Alloy | [211] | 900-1000 °C | Molten Salt (CaO/$CaCl_2$) Electrodes Initial reactant (Ca) | High process complexity, regents required, requires partial artificial atmosphere, high electrical load |

\* Volatile extraction methods (such as water ice evaporation [28, 32]) have been excluded from the current review as they don't cover temperature ranges relevant for metal extraction.

\*\* 1-ethyl-3-methylimidazolium hydrogen sulfate, 1-methylimidazolium hydrogen sulfate, 1-methylpyrrolidinium hydrogen sulfate, 3-[butyl-4-sulfonic acid]-1-methylimidazolium hydrogen sulfate, 3-[butyl-4-sulfonic acid]-1-methylimidazolium triflate [196].

\*\*\* NaCl, sodium 4(2-hydroxyethyl)-1-piperazineethanesulphonic acid, NaOH, $NH_4Cl$, KCl, $NaH_2PO_4·2H_2O$, and trace mineral supplement [208].

\*\*\*\* $H_2SO_4$ processing only targeted Ilmenite as an Fe/$O_2$ source [202]

*Chemical Reduction*

Chemical reduction techniques utilise the preferential reaction of an outside reagent with an oxide to form metal and a secondary oxide. For example, Equation 2 shows the hydrogen reduction of iron monoxide forming Fe metal and water.

$$FeO_{(s)} + H_{2(g)} \rightarrow Fe_{(s)} + H_2O_{(g)} \qquad 2$$

Chemical reduction processes have been used for a long time [1] and are generally well understood. This has led to a relatively simple conversion of these processes for use in extra-terrestrial applications. However, the high TRL of processes that inherently comes from the conversion of well-established industry applications does not infer that their use in an extra-terrestrial setting is favoured.

One of the inherently unavoidable issues with chemical reduction methods for lunar use is the supply chain costs associated with reagent use. Whilst supply chain cost are important in terrestrial industry, very rarely are they so prohibitive as to result in the dismissal of a certain reduction mechanism from a process flowsheet. On the Moon this is not the case. Some processes such as hydrogen reduction can theoretically be operated with a lunar hydrogen source, however this lunar derived hydrogen, as mentioned previously in the case of water, is a valuable resource in its own right. Other processes such as carbothermal reduction, ionic liquid extraction, lithium reduction, acid reduction, calcium reduction, fluorine reduction, and carbochlorination, all currently require terrestrially derived reagents. This issue of supply chain costs is not ignored in the literature, all processes described here have reagent recycling built into the proposed process, however, the question of efficiency at industrial scales requires investigation. Even sub 1% losses in laboratory scale tests will result in prohibitive reagent resupply costs after upscaling. A secondary issue with regent recycling is the inherent process complexity added by this pursuit. The recycling of reagents is achieved through the electrolysis of the secondary oxide described in Equation 2 ($H_2O$, $CO_2$, KF, $Al_2O_3$, etc.). The use of electrolysis for reagent recycling undermines one of the key advantages of chemical reduction methods on Earth, namely the electrical energy requirements.

Several chemical reduction processes (Hydrothermal, Carbothermal, Flourine, and Carbochlorination) also rely on a solid-gas interaction necessitating that the reaction take place in an artificially pressurised atmosphere. Metal reduction using a gaseous reagent cannot be conducted under significantly evacuated conditions due to the pressure of that reagent required for reduction [55]. Similarly, the processes that occur in aqueous environments (Acid reduction, Ionic liquids, and bio-reduction), also require artificial pressurisation to maintain the process reagents in the liquid state. High temperature liquid-phase reactions like those used in metallothermic reduction using lithium, calcium, and aluminium, can theoretically operate in high vacuum conditions however the gradual evaporation of the liquid or slag will result in material losses.

Processes such as the carbochlorination of anorthite ($CaAl_2Si_2O_8$) and the hydrogen reduction of ilmenite ($FeTiO_3$), the latter being more popular in the literature, also suffer from the issue of feedstock availability. The beneficiation of bulk regolith would be required to produce a concentrate stream of the specific mineral feedstock. The beneficiation of the feedstock also warrants scrutiny when considering the effect of gravity and regolith particle size on the reaction kinetics of solid-gas reaction mechanisms. Grill *et al.* [212] analysed the effect of particle sizes on the fluidisation characteristics of lunar regolith simulants, noting that different simulants behaved differently. Two of the simulants (EAC-1 and TUBS-M) had a tendency to experience de-fluidisation, something that was not seen in another simulant material (JSC-1A). Grill *et al.* [212] note that future research using these reactors needs to make sure the simulants used in the research match the desired lunar feedstock to ensure they are designed properly for operation on the Moon. They conclude that the effect of low gravity conditions is an important next step in the validation of the use of a fluidised bed reactor on the lunar surface.

Of the processes categorised here under 'chemical reduction pathways', bio-reduction is one of the more novel and less well studied. The use of biological agents within a mining or metal extraction flowsheet is not common even terrestrially. Whilst summarising NASA's Lunar Regolith Biomining workshop, held in 2007, Dalton *et al.* [213] concluded that "The proposed extraction of O from FeO by cyanobacteria is far beyond the current reach of bioengineering technology". Alternative bacteria such as genetically edited E.Coli strands have since been proposed [209, 210], however these processes still encounter the large issues of precise condition maintenance, extra reagent requirements, and fundamentally, the lack a useful bulk product.

Chemical reduction methods, whilst well understood and easily modifiable for lunar use resulting in high TRL levels, are not ideal metal reduction pathways in a lunar context. With the exception of hydrogen reduction, which only uses Fe-bearing minerals as a feedstock (predominantly ilmenite), and carbothermal reduction using carbon monoxide (the solid-gas reaction variant of the process), the chemical reduction methods are schematically complex processes, that inevitably result in larger supply chain costs over time. There is also a significant lack in the literature addressing known process specific issues such as hydrogen embrittlement of metals and loss through permeation [214]. Whilst some of these processes will be used on the lunar surface in coming years [134], more attention should be given to the understanding and development of processes that are more optimized for off-Earth use.

*Electrochemical Reduction*

Electrochemical reduction processes reduce oxides by passing electrical energy through electrodes and an electrolyte containing or composed of the feedstock material. In an electrolytic process, two separate chemical reactions happen simultaneously, one at each electrode. These reactions are referred to as the anodic reaction (where oxidation occurs) and cathodic reaction (where reduction occurs). Equation 3 is a generalized equation for the electrolysis of a metal oxide that combines the anodic and cathodic reactions into a single equation.

$$2MeO_{(l)} \rightarrow 2Me_{(l)} + O_{2(g)} \qquad 3$$

There are three main categories of electrochemical reduction pathways that have been investigated in the literature. Molten Oxide Electrolysis (MOE), which involves the electrolysis of regolith that has been dissolved in a molten salt electrolyte. Molten Regolith Electrolysis (MRE), which omits the molten salts in favour of using molten regolith as the electrolyte; and Solid Electrolysis, which operates at lower temperatures and uses sintered regolith as the cathode for the cell.

Both MOE and MRE avoid the issues of feedstock particle size and modal composition by working with the regolith material in a molten state. This is very beneficial when considering the arguments presented above in regard to the elimination of the comminution and beneficiation stages of mineral processing. MOE and Solid Electrolysis both still use an electrolyte material that would fall under the category of a required reagent and the detrimental effect of re-supply logistics discussed previously. Solid electrolysis does however have the added benefit of operating at significantly lower temperatures (900 to 950 °C versus the 1300 to 2000 °C of MRE).

Of the three variants, MRE and Solid Electrolysis are preferable. Whilst the use of an electrolyte is not ideal, the added benefit of the lower operating temperature for the Solid Electrolysis has a potential to lower the total energy requirements which may be beneficial in a lunar context. Solid Electrolysis also has the potential to reduce all the oxides present in the regolith creating a metal alloy with an oxygen extraction efficiency of close to 100% [215]. The mixed metal alloy product is not ideal, however, Lomax *et al.* [215] in their demonstration of Solid Electrolysis do suggest to novel approach of designing metal alloys based on prior beneficiation of the feedstock. Power consumption in general for these processes is one of the more potentially limiting issues presented by the task of upscaling to industrial sizes. The major advantage of MRE over Solid electrolysis is that of the lack of an externally sourced electrolyte. MRE uses the regoltih feed material itself as an electrolyte in the reduction process, this comes with the added benefit that solid electrolyte (regolith) can be used to contain the melt as opposed to other insulating materials that are required for MOE and Solid Electrolysis.

Regardless of which electrochemical reduction method is used, all three variants require the use of an inert anode. The pursuit of materials that can be used for such a purpose has been ongoing. Information on the significant amount of work that has been put into the development of 'inert' anode materials for oxide electrolysis can be found elsewhere [163, 197, 216-220]. With current technologies, whilst corrosion rates are extremely low, 7.7mm/year for iridium being considered 'inert' [220], replacement of the anodes would still be required.

Electrochemical reduction has several benefits over chemical reduction methods. Primarily, the process complexity tends to be lower, potentially allowing for much easier automation of the processes. Also, of significance, especially in the case of MRE, the lack of reagents and subsequent re-supply logistics is very appealing. The electrical loads required for these processes are concerning but not insurmountable with increased CAPEX (preferable to the increased operational costs involved with reagent re-supply). Since the reduction is taking place within a molten bath, electrochemical reduction methods do not require beneficiated feedstock and are unaffected by the modal composition and particle size of the feed material. Geochemical composition of the feed material is important in terms of slag acidity/basicity and its effects on cell operation and electrical efficiency. However, these variations are regional and would need to be considered regardless of the reduction method employed. MRE has been identified as one of the primary candidates for ISRU technology by NASA [134].

*Thermal Reduction*

Thermal reduction, historically also referred to as 'Pyrolysis' in the literature, involves the reduction of oxides at high temperatures due to thermal dissociation.

$$2MeO_{(s,l)} \rightarrow 2Me_{(l,g)} + O_{2\,(g)} \qquad\qquad 4$$

This reduction method is not used terrestrially for bulk metal production due to the prohibitive energy requirements of feed material vaporisation. However, in a lunar context, the increased access to higher flux solar radiation and the longer day/night cycle allows for the use of concentrated solar energy as the heat source for prolonged periods of time. The temperature requirements for thermal dissociation vary from 900 to 2700 °C depending on the target oxide and desired reaction rate. Unlike MRE and solid electrolysis, not every oxide can be reduced using this process, some oxides, such as $SiO_2$ and $TiO_2$ form sub-oxides ($SiO$, $Ti_xO_y$) instead of fully reducing under high temperature environments.

The recovery of metal from a thermal dissociation reaction is predicated on the deposition of the metal vapour created in some manner of cold trap. This deposition can be complicated by the back reaction of oxygen gas (formed from the oxide dissociation) with the deposited metal. Based on thermodynamic analysis of oxygen production from a lunar oxide source, Senior [178] concluded that thermal dissociation processes should not exceed 1727 °C as at this temperature the partial pressure of monoatomic oxygen increases over that of oxygen gas in the product. Monoatomic oxygen will react much more readily with metals than will oxygen gas.

Despite its low TRL, thermal dissociation has a number of advantages over traditional metal reduction processes, at least in the context of extra-terrestrial usage. The process requires no reagents or significant wear parts that would require re-supply from terrestrial sources. The majority of the energy requirement for the process can be obtained using concentrated solar flux rather than electrical energy. Whilst increased surface area will increase the kinetics of dissociation, the process theoretically operates on any feedstock regardless of geochemical composition and modality, making it flexible in terms of applicability to other extra-terrestrial feedstocks. And schematically the process is relatively un-complicated. The complexity of recovering usable metal from the cold trap(s) requires significant study. With a fractional deposition method, similar to that used in fractional distillation on Earth, metal vapours can be condensed and recovered individually, but the practicality of this method requires significant research [62, 183].

A further potential benefit to thermal dissociation is the effect of the ambient vacuum on oxide stability as discussed above. The ambient pressure conditions in space, and on the Moon, whilst not helping the kinetics of thermal dissociation, have the potential to significantly reduce the energy requirements of the process. The use of the natural lunar vacuum to replace the pumping equipment usually required for vacuum metallurgical processes on Earth reduces the launch mass of the process. This is the only metal reduction process proposed to date that actively uses the ambient vacuum conditions on the Moon in a beneficial manner. The significant energy reduction requirements for thermal dissociation in vacuum as opposed to at 1 atm has been predicted using thermodynamic modelling in previous work [62]. Similarly, Ellingham Diagrams generated at varying pressures to demonstrate this effect have been presented elsewhere [56]. Whilst the energy reduction for dissociation in vacuum is beneficial, the subsequent slower reaction kinetics require detailed investigation to determine the optimal combination of temperature and pressure for such a process.

It is of note that in the current study, the processes involving selective ionisation [172, 185, 192, 193] have also been included in the category of thermal dissociation. The selective ionisation process differs somewhat to standard thermal dissociation in that the reactor temperature is often well in excess of 7000 °C. The process uses the charge to mass ratio of the ions produced at such high temperatures to separate each atom individually in a charged field [185, 192], in theory resulting in a perfect split of the components of the oxide feed. Study in this field has been theoretical to date.

The significant disadvantages of thermal decomposition as a metal reduction method are the kinetics of the reaction at lower temperatures, and the inability to operate without significant electrical loads during the lunar night. It is also important to note that whilst concentrated solar energy can be used for the primary reduction mechanism, electrical energy will presumably still be required for the cooling of the deposition apparatus. Thermal dissociation as a technology is not advanced enough in TRL to be included in initial ISRU testwork in coming years, however, the combination of flexibility in process feedstock, lack of reagents, relatively low process complexity, and lower electrical energy requirements suggest that this technology, with further research, has the potential to become very prevalent in extra-terrestrial metal production operations in the future.

*Other considerations*

In addition to the primary reduction mechanisms discussed above, there are a number of other challenges that must be taken into account when considering operation on the lunar surface. These issues are not process critical in nature but will need to be considered in detail when more complete end-to-end processing operations are designed. Materials handling in terms of the wear rates of the equipment when moving the abrasive regolith material, the operability of the equipment in dusty and/or electrostatic conditions, the heating characteristics of the regolith material, are all topics that require detailed investigation.

Some studies into this area have already been conducted, for example:

- The flowability of regolith simulant materials in a hopper under various gravitational fields [221]
- The microwave heating characteristics of regolith [222]

- Heat transfer characteristics of lunar regolith and their effect on sintering [223]

There are many more elements of a metal reduction operation that warrant detailed research and development before industrial scale facilities can be operated on the Moon. Whilst the increased access to real lunar regolith and the conditions present on the lunar surface that will be available in coming years will make research in this field significantly easier, there are still very large knowledge gaps that need to be addressed before viable industrial scale process plant designs and subsequent business models can be produced. Some of the more critical challenges will be the equipment design surrounding the processes described here. The design of high vacuum capable seals for reactors that require a pressurised atmosphere. The design of de-gassing chambers for the carbothermal and hydrogen reduction processes that successfully remove all the reagent gas from the reduced feed prior to discharge. Automated tapping mechanisms for processes like MRE that can reliably and safely discharge the metal produced. Reliable methods of heat management in high temperature reactors operating in vacuum conditions.

## Conclusions

The technological advancement and research required for industrial scale ISRU activities targeting minerals other than ice and producing usable metallic products from the lunar regolith is immense. The challenging conditions on the Moon, the low gravity, ultra-high vacuum, intense solar irradiation, electrostatic susceptibility of the regolith, the cryogenic temperatures in shaded regions, and large supply chain costs, represent a challenging suite of considerations that need to be accounted for in the design of mineral processing and metal reduction facilities on the lunar surface.

It has been argued here that the most promising generic flowsheet for a mineral extraction and metal reduction process will: eliminate the comminution circuit, replacing it with basic classification to moderate the particle size of the feed; eliminate beneficiation stages entirely; and utilise schematically simple reduction processes that minimise reagent requirements and can take un-beneficiated, uncomminuted regolith material as a feed source. The metal reduction processes identified as being most applicable to a lunar ISRU/SRU operation are, hydrogen and carbothermal reduction, MRE and solid electrolysis, and vacuum thermal dissociation. Of these, due to the lack of required reagents and subsequent supply chain costs, MRE and vacuum thermal dissociation have the most promise for long term industrial scale implementation. This implementation is subject to concentrated research efforts in these areas to increase TRL levels.

Every aspect of a mineral processing and metal reduction process needs rigorous analysis and a targeted research effort to make the concept of industrial scale ISRU operation a viable option on the Moon. The current work has aimed to provide a thorough review of the relevant conditions on the Moon and their respective effects on potential future mineral processing and metal extraction operations, with the intent to help inform future studies in these areas. It is of note that a number of the considerations portrayed in the current work, especially in terms of the supply chain costs of transport from Earth, are only applicable until large scale resource processing in space can provide the required resources from off-Earth sources, at which time some processes that suffer heavily from the added cost of reagent resupply will be made significantly more viable.

Following this detailed look at mineral processing and metal extraction processes on the lunar surface the following conclusions have been made:

1. Terrestrial processing technologies take advantage of the ambient conditions on Earth; thus, it is incorrect to assume that modification of these technologies for use in space represents the optimal path to ISRU.
2. It is economically beneficial to develop extraction and processing equipment that uses the natural lunar environmental factors advantageously within a process.

3. A remotely controlled or fully automated processing facility is preferable for extra-terrestrial processing operations both economically and when considering human safety.
4. Lunar-based metal extraction operations will favour schematically simple, lightweight processing plants.
5. For beneficiation and reduction processes, reagents should be eliminated wherever possible to minimise supply chain costs.
6. The removal of the comminution and beneficiation circuits, and the development of more robust metal reduction processes that can accommodate un-comminuted and un-beneficiated feed materials has significant merit.
7. The most promising metal reduction processes for industrial scale SRU are MRE and vacuum thermal dissociation.
8. There is a significant amount of research required in all aspects of the field of astrometallurgy to develop and optimise mineral processing and metal extraction technologies for use in space and on the Moon.